\documentclass[12pt,a4paper]{article}

\usepackage{amsmath,amssymb}
\usepackage{graphicx}
\usepackage[nosort]{cite}

\usepackage{tikz}
\usetikzlibrary{decorations.markings}
\usepackage{epsfig}  
\usepackage{enumerate}
\newcommand{\mathsym}[1]{{}}

\usepackage{amsthm,amsbsy,color,multicol}
\usepackage{array,calc}
\usepackage{rotating}
\usepackage{a4wide,bm}
\usepackage[a4paper,text={450pt,650pt},centering]{geometry}
\usepackage{setspace}

\let\oldbfseries=\bfseries
\let\oldmdseries=\mdseries
\let\oldnormalfont=\normalfont
\renewcommand{\bfseries}{\oldbfseries\boldmath}
\renewcommand{\mdseries}{\oldmdseries\unboldmath}
\renewcommand{\normalfont}{\oldnormalfont\unboldmath}

\allowdisplaybreaks[3]

\numberwithin{equation}{section}

\usepackage[font=small,labelfont=bf,width=0.85\textwidth]{caption}

\ifx\hypersetup\sadfkjashdfkxja\newcommand\hypersetup[1]{}\fi

\hypersetup{plainpages=false}
\hypersetup{pdfpagemode=UseNone}
\hypersetup{bookmarksnumbered=true}
\hypersetup{pdfstartview=FitH}
\hypersetup{colorlinks=false}
\hypersetup{citebordercolor={.5 1 .5}}
\hypersetup{urlbordercolor={.5 1 1}}
\hypersetup{linkbordercolor={1 .7 .7}}

\DeclareMathSymbol{\Gamma}{\mathalpha}{letters}{"00}
\DeclareMathSymbol{\Delta}{\mathalpha}{letters}{"01}
\DeclareMathSymbol{\Theta}{\mathalpha}{letters}{"02}
\DeclareMathSymbol{\Lambda}{\mathalpha}{letters}{"03}
\DeclareMathSymbol{\Xi}{\mathalpha}{letters}{"04}
\DeclareMathSymbol{\Pi}{\mathalpha}{letters}{"05}
\DeclareMathSymbol{\Sigma}{\mathalpha}{letters}{"06}
\DeclareMathSymbol{\Upsilon}{\mathalpha}{letters}{"07}
\DeclareMathSymbol{\Phi}{\mathalpha}{letters}{"08}
\DeclareMathSymbol{\Psi}{\mathalpha}{letters}{"09}
\DeclareMathSymbol{\Omega}{\mathalpha}{letters}{"0A}

\newcommand{\gen}[1]{\mathrm{#1}}

\newcommand{\dd}{\mathrm{d}}
\newcommand{\ii}{\mathrm{i}}

\ifx\genfrac\sdflkaj\else\fi

\newcommand{\ket}[1]{\left|#1\right\rangle}      
\newcommand{\bra}[1]{\left\langle #1\right|}     

\newcommand{\alg}[1]{\mathfrak{#1}}

\newcommand{\beq}{\begin{equation}}
\newcommand{\eeq}{\end{equation}}

\def\[{\begin{equation}}
\def\]{\end{equation}}
\def\<{\begin{eqnarray}}
\def\>{\end{eqnarray}}

\makeatletter
\def\mr@ignsp#1 {\ifx\:#1\@empty\else #1\expandafter\mr@ignsp\fi}%
\newcommand{\multiref}[1]{\begingroup
\xdef\mr@no@sparg{\expandafter\mr@ignsp#1 \: }%
\def\mr@comma{}%
\@for\mr@refs:=\mr@no@sparg\do{\mr@comma\def\mr@comma{,}\ref{\mr@refs}}%
\endgroup}
\makeatother

\newcommand{\hypref}[2]{\ifx\href\asklfhas #2\else\href{#1}{#2}\fi}
\newcommand{\Secref}[1]{Section~\multiref{#1}}

\newcommand{\Appref}[1]{Appendix~\multiref{#1}}

\renewcommand{\eqref}[1]{(\multiref{#1})}

\makeatletter
\newlength{\apb@width}
\newcommand{\autoparbox}[2][c]{\settowidth{\apb@width}{#2}\parbox[#1]{\apb@width}{#2}}

\makeatother

\ifx\href\asklfhas\newcommand{\href}[2]{#2}\fi

\begin{document}

\renewcommand{\thefootnote}{\fnsymbol{footnote}}
\thispagestyle{empty}
\begin{flushright}\footnotesize
\end{flushright}
\vspace{1cm}

\begin{center}%
{\Large\bfseries%
\hypersetup{pdftitle={Refined functional relations for the elliptic SOS model}}%
Refined functional relations \\ for the elliptic SOS model%
\par} \vspace{2cm}%

\textsc{W. Galleas}\vspace{5mm}%
\hypersetup{pdfauthor={Wellington Galleas}}%

\textit{ARC Centre of Excellence for the Mathematics  \\ and Statistics of Complex Systems, \\%
The University of Melbourne\\%
VIC 3010, Australia}\vspace{3mm}%

\verb+wgalleas@unimelb.edu.au+ %

\par\vspace{3cm}

\textbf{Abstract}\vspace{7mm}

\begin{minipage}{12.7cm}
In this work we refine the method of \cite{Galleas_2011} and obtain a novel kind of
functional equation determining the partition function of the elliptic SOS model
with domain wall boundaries. This functional relation arises from the dynamical Yang-Baxter relation and
its solution is given in terms of multiple contour integrals.

\hypersetup{pdfkeywords={dynamical Yang-Baxter equation, domain wall boundaries, functional equations}}%
\hypersetup{pdfsubject={}}%
\end{minipage}
\vskip 2cm
{\small PACS numbers:  05.50+q, 02.30.IK}
\vskip 0.1cm
{\small Keywords: Dynamical Yang-Baxter equation, Functional relations, \\ Domain wall boundaries}
\vskip 2cm
{\small August 2012}

\end{center}

\newpage
\renewcommand{\thefootnote}{\arabic{footnote}}
\setcounter{footnote}{0}

\tableofcontents

\section{Introduction}
\label{sec:intro}

Face models or Solid-on-Solid (SOS) models of statistical mechanics were introduced
by Baxter in the process of solving the eight-vertex model with periodic boundary 
conditions \cite{Baxter_1973}. The Boltzmann weights of Baxter's eight-vertex model are
parameterised by elliptic functions and this feature is intrinsically connected
with the requirement that the model statistical weights satisfy the Yang-Baxter
equation \cite{Baxter_1971, Baxter_1972}.
The elliptic nature of the eight-vertex model Boltzmann weights is naturally
transported to the corresponding SOS model and a new continuous parameter emerges in the course of
Baxter's vertex-face transformation \cite{Baxter_1973}.
We shall refer to this new parameter as dynamical parameter \cite{Baxter_1973}
and its implications for the analytic theory of the eight-vertex model have been
discussed in \cite{Bazhanov_2007}. Besides the emergence of this new parameter, the
resulting statistical weights no longer satisfy the standard Yang-Baxter equation but its dynamical version introduced in
\cite{Gervais_1984} and subsequently considered by Felder \cite{Felder_94, Felder_1994, Felder_1996}
as the quantised form of a modified classical Yang-Baxter equation \cite{Bernard1_1988, Bernard2_1988}. 

In the same fashion as Drinfeld-Jimbo quantum groups \cite{drinfeld, jimbo1, jimbo2, jimbo3}
provide the algebraic structure underlying the solutions of the Yang-Baxter equation, 
the so called elliptic quantum groups introduced in \cite{Felder_94, Felder_1994} accommodate 
the solutions of the dynamical Yang-Baxter equation. In this work we shall restrict ourselves to the
SOS model built out of the solution of the dynamical Yang-Baxter equation associated with the elliptic
quantum group $E_{\tau , \gamma}[\alg{sl}_2]$. As far as the boundary conditions are concerned, we shall
consider the case of domain wall boundaries firstly introduced by Korepin in the context of vertex models \cite{Korepin82}
and subsequently extended for SOS models in \cite{Galleas_2011, Rosengren_2008, Pakuliak_2008, WengLi_2009}.

In contrast to the case with periodic boundary conditions, the partition function of vertex and SOS models
with domain wall boundaries can be exactly computed without relying on solutions of Bethe ansatz equations.
Interestingly enough, the exact solution of the six-vertex model with domain wall boundaries \cite{Izergin87}
revealed that the model free-energy differs from the case with periodic boundary conditions \cite{Justin_2000}.
This unusual dependence of bulk thermodynamic properties with boundary conditions has also been observed for the 
elliptic SOS model when the anisotropy parameter assumes a particular value \cite{Rosengren_2011}.
For general values of the anisotropy parameter this study still poses as an open problem, probably
due to the lack of suitable expressions for the partition function allowing to compute
physical properties in the thermodynamic limit. In searching for alternative representations for this
partition function, which might render the analysis of the thermodynamic limit feasible, 
we have obtained in \cite{Galleas_2011} a multiple integral formula for the partition function of the trigonometric SOS model with domain wall
boundaries. This case consists of a particular limit of a more general elliptic model, the limit where
elliptic theta-functions degenerate into trigonometric functions, and here we refine and generalise
the method of \cite{Galleas_2011} for the general elliptic case.

This paper is planned as follows. In the \Secref{sec:dyn} we give a brief description
of SOS models with domain wall boundaries in terms of the generators of Felder's 
dynamical Yang-Baxter relations. The conventions employed here are basically the ones
already discussed in  \cite{Galleas_2011}. In the \Secref{sec:fun} we demonstrate
how the dynamical Yang-Baxter relations can be explored in order to obtain a functional equation 
determining the model partition function. This functional equation is solved in \Secref{sec:partition}
and concluding remarks are discussed in \Secref{sec:conclusion}. Technical details required throughout
this paper are presented in the Appendices.

\section{Operatorial description of the SOS model}
\label{sec:dyn}

Partition functions of two-dimensional lattice models can be described in terms of
operators representing the allowed configurations of the lattice.
This feature goes back to Kramers and Wannier transfer matrix technique \cite{Kramers_1941a,Kramers_1941b}
and it has found several important generalisations \cite{Baxter_book}. 
Remarkably, when the statistical weights of the model satisfy the Yang-Baxter equation \cite{Baxter_1971} or its
dynamical counterpart \cite{Felder_94, Felder_1994, Felder_1996}, we not only have an operatorial description
of the model but also an algebra governing its operators for any size of the lattice. 
In what follows we shall recall the conventions discussed in \cite{Galleas_2011} which consist of an 
extension of the ones  given in \cite{Korepin82} for the six-vertex model. 

\paragraph{Dynamical Yang-Baxter equation.} Following \cite{Felder_1994, Felder_1996} we encode the statistical
weights of our elliptic SOS model on a matrix $\mathcal{R} \in \mbox{End}(\mathbb{V} \otimes \mathbb{V})$ 
with $\mathbb{V} \cong \mathbb{C}^{2}$. For variables $\lambda_i , \gamma , \theta \in \mathbb{C}$, this matrix 
$\mathcal{R}$ satisfies the dynamical Yang-Baxter equation,
\<
\label{yb}
\mathcal{R}_{12}(\lambda_1 - \lambda_2, \theta - \gamma \hat{h}_3) \mathcal{R}_{13}(\lambda_1 - \lambda_3, \theta) \mathcal{R}_{23}(\lambda_2 - \lambda_3, \theta - \gamma \hat{h}_1 ) = \nonumber \\
\mathcal{R}_{23}(\lambda_2 - \lambda_3, \theta) \mathcal{R}_{13}(\lambda_1 - \lambda_3, \theta - \gamma \hat{h}_2) \mathcal{R}_{12}(\lambda_1 - \lambda_2,\theta) \; ,
\>
where $\hat{h} = \mbox{diag}(1,-1)$. The Eq. (\ref{yb}) is defined in $\mbox{End} (\mathbb{V}_1  \otimes \mathbb{V}_2  \otimes \mathbb{V}_3 )$ and
the action of $\mathcal{R}_{12}(\lambda, \theta - \gamma \hat{h}_3 )$ on the basis vector $v_1 \otimes v_2 \otimes v_3$ is understood as
\[
\left[ \mathcal{R}(\lambda, \theta - \gamma h) v_1 \otimes v_2 \right] \otimes v_3 \; ,
\]
where $h$ is a scalar denoting a particular eigenvalue of $\hat{h}$, i.e. $\hat{h}_i v_i = h v_i$.
 
\paragraph{Definition.} Let $\tau$ be a complex number such that $\mbox{Im}(\tau) > 0$ and write $p = e^{\ii \pi \tau}$
so that $|p|<1$. For $\lambda \in \mathbb{C}$ we define the elliptic function $f$ with nome $p$ as
\[
\label{ff}
f(\lambda) = \frac{1}{2} \sum_{n=-\infty}^{\infty} (-1)^{n - \frac{1}{2}} p^{(n+\frac{1}{2})^2} e^{-(2n+1)\lambda} \; .
\]
The function $f(\lambda)$ corresponds to the Jacobi theta-function $\Theta_1(\ii \lambda,\tau)/2$ \cite{Watson}
and in the \Appref{sec:elliptic} we have collected the properties of $f$ required through this work.

\bigskip 

The equation (\ref{yb}) has been considered in \cite{Felder_94, Felder_1994} and its solution reads
\[
\label{rmat}
\mathcal{R} (\lambda, \theta) = \left( \begin{matrix}
a_{+}(\lambda, \theta) & 0 & 0 & 0 \cr 
0 & b_{+}(\lambda, \theta) & c_{+}(\lambda, \theta) & 0 \cr
0 & c_{-}(\lambda, \theta) & b_{-}(\lambda, \theta) & 0 \cr
0 & 0 & 0 & a_{-}(\lambda, \theta) \end{matrix} \right) 
\]
with non-null entries 
\<
\label{bw}
a_{\pm}(\lambda, \theta) &=& f(\lambda + \gamma) \nonumber \\
b_{\pm}(\lambda, \theta) &=& f(\lambda) \frac{f(\theta \mp \gamma)}{f(\theta)} \nonumber  \\
c_{\pm}(\lambda, \theta) &=& f(\gamma) \frac{f(\theta \mp \lambda)}{f(\theta)} \;\; .
\>
Although we shall not make explicit use of it, we remark here that the algebraic structure 
underlying (\ref{rmat}) is the elliptic quantum group $E_{\tau,\gamma}[\alg{sl}_2]$ \cite{Felder_94}.

\paragraph{Dynamical monodromy matrix.} Let $\hat{\theta}_i$ be the operator valued parameter
\[
\hat{\theta}_i = \theta - \gamma \displaystyle \sum_{k=i+1}^L \hat{h}_{k}
\]
and consider the following ordered product of dynamical $\mathcal{R}$-matrices,
\[
\label{mono}
\mathcal{T}_{a} (\lambda, \theta) = \mathop{\overrightarrow\prod}\limits_{1 \le i \le L } \mathcal{R}_{a i}(\lambda - \mu_i, \hat{\theta}_i) \; ,
\]
living in the tensor product space $\mathbb{V}_a \otimes \mathbb{V}_1 \otimes \dots \otimes \mathbb{V}_L$.
We shall refer to $\mathcal{T}_{a} (\lambda, \theta)$ as dynamical monodromy matrix or simply monodromy matrix.
Since the dynamical $\mathcal{R}$-matrix (\ref{rmat}) satisfy the weight-zero condition 
$[ \mathcal{R}_{ab} (\lambda, \theta), \hat{h}_a + \hat{h}_b ] = 0$, one can show that (\ref{mono}) obeys the relation
\<
\label{DYB}
\mathcal{R}_{ab}(\lambda_1 - \lambda_2, \theta - \gamma \gen{H}) \mathcal{T}_a (\lambda_1, \theta) \mathcal{T}_b (\lambda_2, \theta - \gamma \hat{h}_a) =  
\mathcal{T}_b (\lambda_2, \theta) \mathcal{T}_a (\lambda_1, \theta - \gamma \hat{h}_b ) \mathcal{R}_{ab}(\lambda_1 - \lambda_2, \theta) \nonumber \\
\>
with $\gen{H} = \displaystyle \sum_{k=1}^L \hat{h}_{k}$. Here we are considering $\mathbb{V} \cong \mathbb{C}^2$
and the dynamical monodromy matrix can be recast in the form
\[
\label{abcd}
\mathcal{T}_{a} (\lambda, \theta) = \left( \begin{matrix}
A(\lambda, \theta) & B(\lambda, \theta) \cr
C(\lambda, \theta) & D(\lambda, \theta) \end{matrix} \right) 
\]
whose entries are then defined on $\mathbb{V}_1 \otimes \dots \otimes \mathbb{V}_L$. The formula
(\ref{DYB}) encodes commutation relations for the entries of (\ref{abcd}) which shall be referred
to as dynamical Yang-Baxter relations.

\paragraph{Domain wall boundaries.} The partition function of the elliptic SOS model with domain wall boundaries
can be written in terms of entries of (\ref{abcd}) as described in \cite{Galleas_2011}. More precisely, 
the elliptic SOS model partition function $Z_{\theta}$ is given by the expected value
\[
\label{pft}
Z_{\theta} = \bra{\bar{0}} \mathop{\overrightarrow\prod}\limits_{1 \le j \le L } B(\lambda_j, \theta + j \gamma) \ket{0} 
\]
where
\[
\label{states1}
\ket{0} = \bigotimes_{i=1}^{L} \left( \begin{matrix}
1 \cr
0 \end{matrix} \right) \qquad \qquad \mbox{and} \qquad \qquad
\ket{\bar{0}} = \bigotimes_{i=1}^{L} \left( \begin{matrix}
0 \cr
1 \end{matrix} \right) \; .
\]
In the next section we shall demonstrate how the dynamical Yang-Baxter relations can be employed
to produce a functional equation determining $Z_{\theta}$.

\section{Functional relations}
\label{sec:fun}

The relation (\ref{DYB}) encodes commutation rules for the operators $A(\lambda, \theta)$, $B(\lambda, \theta)$,
$C(\lambda, \theta)$ and $D(\lambda, \theta)$ once the structure (\ref{abcd}) is considered.  
Out of the sixteen relations contained in (\ref{DYB}), we will make use of only two of them
in order to derive a functional equation describing the partition function (\ref{pft}). More precisely, 
the required relations are simply:
\<
\label{alg}
B(\lambda_1, \theta) B(\lambda_2, \theta +\gamma) &=& B(\lambda_2, \theta) B(\lambda_1, \theta +\gamma) \nonumber \\
A(\lambda_1, \theta + \gamma) B(\lambda_2, \theta) &=& \frac{f(\lambda_2 - \lambda_1 +\gamma)}{f(\lambda_2 - \lambda_1)} \frac{f(\theta +\gamma)}{f(\theta + 2 \gamma)} B(\lambda_2, \theta + \gamma) A(\lambda_1, \theta + 2\gamma) \nonumber \\
&-& \frac{f(\theta + \gamma - \lambda_2 + \lambda_1)}{f(\lambda_2 - \lambda_1)} \frac{f(\gamma)}{f(\theta + 2\gamma)} B(\lambda_1, \theta + \gamma) A(\lambda_2, \theta + 2\gamma)  \; . \nonumber \\
\>
In addition to that, the weight-zero condition satisfied by (\ref{rmat}) associated with the
definition (\ref{mono}) allows us to compute the action of $A(\lambda, \theta)$ on the states
$\ket{0}$ and $\ket{\bar{0}}$ defined in (\ref{states1}). The vectors $\ket{0}$ and $\ket{\bar{0}}$
are respectively the $\alg{sl}_2$ highest and lowest weight states and 
from (\ref{rmat}) and (\ref{mono}) we readily obtain
\<
\label{action}
A(\lambda, \theta) \ket{0} &=& \prod_{j=1}^{L} f(\lambda - \mu_j + \gamma) \; \ket{0}  \nonumber \\
\bra{\bar{0}} A(\lambda, \theta) &=& \frac{f(\theta - \gamma)}{f(\theta + (L-1)\gamma)} \prod_{j=1}^{L} f(\lambda - \mu_j) \; \bra{\bar{0}} \; .
\>

\paragraph{The framework.} In order to explore the relations (\ref{alg}) and (\ref{action})
we shall consider the quantity
\[
\label{source}
\bra{\bar{0}} A(\lambda_0, \theta + \gamma) Y_{\theta - \gamma} (\lambda_1 , \dots , \lambda_L) \ket{0} \; ,
\]
where $Y_{\theta} (\lambda_1 , \dots , \lambda_L) = \mathop{\overrightarrow\prod}\limits_{1 \le j \le L } B(\lambda_j, \theta + j\gamma)$, 
computed in two different ways. One of them only makes use of the properties (\ref{action}) arising from 
the $\alg{sl}_2$ highest weight representation theory, while the second way employ the dynamical Yang-Baxter
relations (\ref{alg}) in addition to (\ref{action}). For instance, we can compute the term $\bra{\bar{0}} A(\lambda_0, \theta + \gamma)$ using solely (\ref{action})
to find that (\ref{source}) is proportional to $Z_{\theta - \gamma}(\lambda_1 , \dots , \lambda_L)$.
On the other hand, we could have firstly examined the quantity $A(\lambda_0, \theta + \gamma) Y_{\theta - \gamma} (\lambda_1 , \dots , \lambda_L) \ket{0}$.
For that we employ the relations (\ref{alg}) to move the operator $A(\lambda_0, \theta + \gamma)$ through 
the string of operators $B(\lambda_j, \theta + (j-1)\gamma)$ and then consider the action of the resulting
operator $A$ on the vector $\ket{0}$. Exacting this procedure, we repeatedly apply (\ref{alg})
together with the addition rule (\ref{add}) in order to show that
\<
\label{AY}
&& A(\lambda_0, \theta + \gamma) Y_{\theta - \gamma} (\lambda_1 , \dots , \lambda_L) = \nonumber \\
&& \frac{f(\theta + \gamma)}{f(\theta + (L+1) \gamma)} \prod_{j=1}^{L} \frac{f(\lambda_j - \lambda_0 + \gamma)}{f(\lambda_j - \lambda_0)} Y_{\theta} (\lambda_1 , \dots , \lambda_L) A(\lambda_0 , \theta + (L+1)\gamma) \nonumber \\
&& - \sum_{i=1}^{L} \frac{f(\theta + \gamma - \lambda_i + \lambda_0)}{f(\theta + (L+1)\gamma)} \frac{f(\gamma)}{f(\lambda_i - \lambda_0)} \prod_{\stackrel{j=1}{j \neq i}}^{L} \frac{f(\lambda_j - \lambda_i + \gamma)}{f(\lambda_j - \lambda_i)} \times \nonumber \\
&& \qquad \quad Y_{\theta} (\lambda_0 , \lambda_1 , \dots , \lambda_{i-1} , \lambda_{i+1} , \dots , \lambda_L) A(\lambda_i , \theta + (L+1)\gamma) \; .
\>
Then we use (\ref{action}) to compute the action of the operators $A(\lambda_i , \theta + (L+1)\gamma)$ appearing on the RHS of (\ref{AY})
on the vector $\ket{0}$. Thus the combination of (\ref{AY}) and (\ref{action}) allows us to write the quantity (\ref{source})
as a linear combination of terms $Z_{\theta}$ depending on the set of $L+1$ variables
$\{ \lambda_0 , \lambda_1 , \dots , \lambda_L \}$  where only $L$ variables are taken at a time. 

\paragraph{Functional equation.} Taking into account the above discussion, we can see that the consistency
between the $\alg{sl}_2$ highest weight representation theory, manifested in the relations (\ref{action}),
and the dynamical Yang-Baxter relations implies the functional equation
\<
\label{FZ}
M_0 \; Z_{\theta - \gamma}(\lambda_1, \dots , \lambda_L) + \sum_{i=0}^{L} N_i \; Z_{\theta}(\lambda_0, \dots , \lambda_{i-1}, \lambda_{i+1}, \dots, \lambda_L) = 0  \; ,  
\>
with coefficients given by
\<
\label{coeff}
M_0 &=& \frac{f(\theta)}{f(\theta + L\gamma)} \prod_{j=1}^{L} f(\lambda_0 - \mu_j) \nonumber \\
N_0 &=& - \frac{f(\theta + \gamma)}{f(\theta + (L+1)\gamma)} \prod_{j=1}^{L} f(\lambda_0 - \mu_j + \gamma) \prod_{j=1}^{L} \frac{f(\lambda_j - \lambda_0 + \gamma)}{f(\lambda_j - \lambda_0)} \nonumber \\
N_i &=& \frac{f(\theta + \gamma + \lambda_0 - \lambda_i)}{f(\theta + (L+1)\gamma)} \frac{f(\gamma)}{f(\lambda_i - \lambda_0)} \prod_{j=1}^{L} f(\lambda_i - \mu_j + \gamma) \prod_{\stackrel{j=1}{\neq i}}^{L} \frac{f(\lambda_j - \lambda_i + \gamma)}{f(\lambda_j - \lambda_i)} \nonumber \\
& & \qquad \qquad \qquad \qquad \qquad \qquad \qquad \qquad \qquad \qquad \qquad \qquad \qquad \quad  i=1, \dots , L \; .   \nonumber \\ 
\>
Some remarks are in order at this stage. Although the partition function considered here reduces 
to the one studied in \cite{Galleas_2011} when the elliptic theta-function $f$ degenerate into 
a trigonometric function, the functional equation (\ref{FZ}) still differs significantly from 
the one obtained in \cite{Galleas_2011}. For instance, (\ref{FZ}) is a functional equation also over
the variable $\theta$ and even in the limit $\theta \rightarrow \infty$, where $Z_{\theta - \gamma}$ and $Z_{\theta}$ coincide,
we still would be left with a functional equation different from the one presented in \cite{Galleas10}.
This divergence is due to the fact that here we have started our analysis with the quantity (\ref{source}) instead of 
$\bra{\bar{0}} C(\lambda_0, \theta + \gamma) \prod_{j=1}^{L+1} B(\lambda_j, \theta + (j-1)\gamma) \ket{0}$ as employed
in the works \cite{Galleas_2011} and \cite{Galleas10}. This different starting point allows us to obtain a simpler functional
equation whose solution will be discussed in the next section.

\section{The partition function}
\label{sec:partition}

This section is concerned with solving the functional relation (\ref{FZ}). The method
we shall employ is essentially the one described in \cite{Galleas_2011} which exploits special
zeroes of $Z_{\theta}$ to produce a separation of variables. 
Some structural properties of (\ref{FZ}) will be of utility to help us identifying the elements 
required to solve this functional equation. For instance, the partition function $Z_{\theta}$ is a function of
two sets of variables, i.e. $\{ \lambda_1 , \dots , \lambda_L \}$ and $\{ \mu_1 , \dots , \mu_L \}$,
in addition to the parameters $\gamma$, $\theta$ and the elliptic nome $p$. In our framework, however, the 
set of variables $\{ \mu_1 , \dots , \mu_L \}$ can also be regarded as parameters while $\theta$ is promoted to a variable. 
This follows from the fact that (\ref{FZ}) is an equation not only over variables $\lambda_j$ but also $\theta$. 

With this in mind we can see that (\ref{FZ}) is a homogeneous equation in the sense that $\alpha Z_{\theta}$ is a solution 
if so is $Z_{\theta}$ and $\alpha$ is independent of $\lambda_j$ and $\theta$. This property implies that 
the equation (\ref{FZ}) will be able to determine the partition function up to an overall multiplicative
factor independent of $\lambda_j$ and $\theta$ at most. Thus the complete determination of $Z_{\theta}$ will require
that we are able to compute it for a particular value of $\lambda_j$ and $\theta$ in order to determine this overall
factor. Any point on the $(\lambda_j , \theta)$-plane would serve our need and we can choose
it such that the evaluation of $Z_{\theta}$ is as simple as possible. As demonstrated in the \Appref{sec:asymp}, the evaluation of
$Z_{\theta}$ in the limit $(\lambda_j , \theta ) \rightarrow \infty$ can be performed in the same lines of \cite{Galleas10}.
Moreover, the equation (\ref{FZ}) is linear which raises the issue of uniqueness of the solution since linear combinations
of particular solutions also solve (\ref{FZ}). Similarly to the case considered in \cite{Galleas_2011}, we will
see that the location of zeroes of $Z_{\theta}$ will select the appropriate solution uniquely.
The asymptotic behaviour of $Z_{\theta}$ is obtained in the Appendix A while the characterisation of the partition function
in terms of its zeroes is discussed in the Appendix B. These properties are summarised as follows.

\paragraph{Asymptotic behaviour.} In the limit $(\lambda_j , \theta ) \rightarrow \infty$ the partition function (\ref{pft}) behaves as
\<
\label{asy}
&& Z_{\theta} (\lambda_1 , \dots , \lambda_L) \sim \nonumber \\
&& \frac{f(\gamma)^L}{2^{L(L-1)}} \sum_{n_1^{(1)} = - \infty}^{\infty} \dots \sum_{n_{L-1}^{(1)} = - \infty}^{\infty} \dots \sum_{n_1^{(L)} = - \infty}^{\infty} \dots \sum_{n_{L-1}^{(L)} = - \infty}^{\infty} (-1)^{\sum_{a=1}^{L} \sum_{i=1}^{L-1} n_i^{(a)} - \frac{L(L-1)}{2}} \nonumber \\
&& \qquad \qquad  \prod_{a=1}^{L} \prod_{i=1}^{L-1} p_{n_i^{(a)}} q_{n_i^{(a)}} e_{n_i^{(a)}}^{\lambda_a - \mu_{i}^{(a)}} 
\; \sum_{\sigma \in \mathcal{S}_L} \prod_{(a,b) \in I_{\sigma}} (q_{n_{b-1}^{(a)}} q_{n_{a}^{(b)}})^{-1} \; ,
\>
where $e_n = e^{-(2n + 1)}$, $p_n = p^{(n + \frac{1}{2})^2}$, $q_n = e_n^{\gamma}$ and $\mu^{(a)} = \{ \mu_i : i \neq a \}$.
Here $\mathcal{S}_L$ denotes the group of permutations of $L$ objects and $\sigma = \sigma(1) \dots \sigma(L)$
stands for a given permutation. The set of inversion vertices for a given $\sigma$ is denoted 
by $I_{\sigma}$.

\medskip
\paragraph{Higher order theta-function.} The partition function $Z_{\theta}$ is a theta-function of order $L$ and norm $t_i$
in each one of its variables $\lambda_i$ separately. That is to say there exist constants $C$ and $\xi_j^{(i)}$
satisfying $\xi_1^{(i)} + \dots + \xi_L^{(i)} = t_i$ such that
\<
\label{ells}
Z_{\theta} = C \prod_{j=1}^{L} f(\lambda_i - \xi_j^{(i)}) \; .
\>
Although an explicit expression for the norm $t_i$ shall not be required, unveiling special zeroes $\xi_j^{(i)}$
for a particular specialisation of variables will be an important step for solving (\ref{FZ}).

\bigskip
Now we shall proceed with the analysis of (\ref{FZ}) in the lines of \cite{Galleas_2011}. For that we look
for special values of the variables $\lambda_j$ such that particular zeroes of $Z_{\theta}$
can be identified.

\paragraph{Special zeroes.} The coefficients $M_0$ and $N_i$ given in (\ref{coeff})
exhibit a factorised form and due to that identifying their zeroes is a simple task. For instance,
when $\lambda_0 = \mu_1$ and $\lambda_1 = \mu_1 - \gamma$ we find that $M_0 = N_0 = N_1 = 0$. 
Next we set $\lambda_j = \lambda_{j+1} - \gamma$ successively for $j \in [2, L-1]$ and collect the result at each step.
At the last step we find
\[
\label{vanish}
N_2 \;  Z_{\theta} (\mu_1 , \mu_1 - \gamma , \lambda_L - (L-3)\gamma, \lambda_L - (L-4)\gamma, \dots , \lambda_L) = 0 \; ,
\]
and since $N_2$ is different from zero we can conclude that the vanishing of (\ref{vanish}) is due to $Z_{\theta}$. This result can
now be substituted back into the previous steps leading to (\ref{vanish}). By doing so we find the more general
vanishing condition, namely $Z_{\theta} (\mu_1 , \mu_1 - \gamma , \lambda_3, \dots , \lambda_L) = 0$, for general values of the
variables $\lambda_j$ with $j \in [3, L]$. This process can also be performed starting with variables
$\lambda_0 = \mu_1$ and $\lambda_j = \mu_1 - \gamma$ for any $j \in [1,L]$, which allows us to conclude that
$Z_{\theta} (\mu_1 , \dots , \mu_1 - \gamma , \dots) = 0$. 

\bigskip

\paragraph{Building up the solution.} The zeroes of $Z_{\theta}$ above unveiled have a special appeal since we are
interested in the solution of (\ref{FZ}) consisting of a higher order theta-function (\ref{ells}). Taking that into account,
those special zeroes imply that
\[
\label{red}
Z_{\theta} (\mu_1 , \lambda_2 , \dots , \lambda_L) = \prod_{j=2}^{L} f(\lambda_j - \mu_1 + \gamma) \; V_{\theta} (\lambda_2 , \dots , \lambda_L) \; ,
\]
where $V_{\theta}$ is also a theta-function but of order $L-1$ in each one of its variables.
Next we set $\lambda_0 = \mu_1$ in the Eq. (\ref{FZ}) and substitute the expression (\ref{red})
into it. The resulting equation can then be solved for $Z_{\theta} (\lambda_1 , \dots , \lambda_L)$
yielding the formula
\[
\label{red1}
Z_{\theta} (\lambda_1 , \dots , \lambda_L) = \sum_{i=1}^{L} m_i \;  V_{\theta} (\dots, \lambda_{i-1}, \lambda_{i+1}, \dots)
\]
with coefficients
\<
\label{mi}
m_i = \frac{f(\theta + \gamma + \mu_1 - \lambda_i)}{f(\theta + \gamma)} \prod_{j=2}^{L} \frac{f(\lambda_i - \mu_j + \gamma)}{f(\mu_1 - \mu_j + \gamma)} \prod_{\stackrel{j=1}{\neq i}}^{L} f(\lambda_j - \mu_1) \frac{f(\lambda_j - \lambda_i + \gamma)}{f(\lambda_j - \lambda_i)} \; .
\>
We then substitute the formula (\ref{red1}) back into the original equation (\ref{FZ}) and set $\lambda_L = \mu_1$. After eliminating 
an overall factor we are left with the equation
\[
\label{FZr}
P_0 \; V_{\theta - \gamma}(\lambda_1, \dots , \lambda_{L-1}) + \sum_{i=0}^{L-1} Q_i \; V_{\theta}(\lambda_0, \dots , \lambda_{i-1}, \lambda_{i+1}, \dots, \lambda_{L-1}) = 0 \; ,   
\]
where the coefficients $P_0$ and $Q_i$ correspond respectively to the coefficients $M_0$ and $N_i$ given
in (\ref{coeff}) under the mapping $L \rightarrow L-1$, $\theta \rightarrow \theta + \gamma$ 
and $\mu_i \rightarrow \mu_{i+1}$. Thus the function $V_{\theta}$ obeys essentially the same equation
as the partition function $Z_{\theta}$ but for a square lattice of dimensions $(L-1) \times (L-1)$. 
Now since $V_{\theta}$ is also a theta-function, this procedure can be repeatedly carried out until we reach the equation
for $L=1$. The solution of (\ref{FZ}) for $L=1$ can be found in the \Appref{sec:sol1} and gathering our results 
we obtain the following solution for general $L$,
\[
\label{solL}
Z_{\theta}(\lambda_1 , \dots , \lambda_L) = \sum_{\sigma \in \mathcal{S}_L} F_{\sigma(1) \dots \sigma(L)} 
\]
where
\begin{align}
\label{solF}
& F_{\sigma(1) \dots \sigma(L)} = \nonumber \\
& \frac{\Omega_L}{\prod_{k=2}^{L} f(\mu_1 - \mu_k + \gamma)} \prod_{n=1}^{L} \frac{f(\theta + n \gamma - \lambda_{\sigma(n)} + \mu_n)}{f(\theta + n \gamma)} \prod_{j>n}^{L} f(\lambda_{\sigma(n)} - \mu_j + \gamma) \prod_{j<n}^{L} f(\lambda_{\sigma(n)} - \mu_j) \nonumber \\
& \times \; \prod_{m>n}^{L} \frac{f(\lambda_{\sigma(m)} - \lambda_{\sigma(n)} + \gamma)}{f(\lambda_{\sigma(m)} - \lambda_{\sigma(n)})} \; .
\end{align}
The overall factor $\Omega_L$ arises from the homogeneity of (\ref{FZ}) as previously discussed, and from (\ref{asy}) we 
obtain $\Omega_L = f(\gamma)^L \prod_{k=2}^{L} f(\mu_1 - \mu_k + \gamma)$. It is important to remark here
that this partition function has also been considered in \cite{Rosengren_2008, Pakuliak_2008, WengLi_2009}
where a similar but still different expression for $F_{\sigma(1) \dots \sigma(L)}$ has been found.

\paragraph{Multiple integral formula.} The partition function $Z_{\theta}$ can be represented by a multiple contour
integral as follows. The function $V_{\theta}$ in the formula (\ref{red1}) is essentially the partition
function for a lattice of size $(L-1)\times (L-1)$ and modified parameters. In fact, the decomposition of
$Z_{\theta}$ in terms of $V_{\theta}$ as described by (\ref{red1}) can be thought of as a separation of 
variables. Moreover, we shall see that the prescription given by (\ref{red1}) can be mimicked by the Cauchy like integral
\<
\label{ansatz}
Z_{\theta} (\lambda_1, \dots , \lambda_L) = \oint \dots \oint \frac{H (w_1 , \dots , w_L )}{\prod_{i,j=1}^{L} f(w_i - \lambda_j)} \prod_{j=1}^{L} \frac{\dd w_j}{2 \ii \pi} , 
\>
with integration contours enclosing solely the zeroes of $f$ when $w_i \rightarrow \lambda_j$. 
Also we shall assume that $H ( w_1 , \dots , w_L )$ has no poles inside the integration contour.
Under those assumptions the formula (\ref{ansatz}) can be for instance, integrated over the variable
$w_1$, and by doing so we obtain the relation 
\<
\label{ansatz1}
Z_{\theta} (\lambda_1, \dots , \lambda_L) = \sum_{i=1}^{L} f'(0)^{-1} \oint \dots \oint \frac{\left. H (w_1 , \dots , w_L ) \right|_{w_1 = \lambda_i}}{\prod_{j \neq i}^{L} f(\lambda_i - \lambda_j) \prod_{j=2}^{L} f(w_j - \lambda_i)}  \times \nonumber \\
\qquad \qquad \qquad \qquad \qquad \qquad \frac{1}{\prod_{k=2}^{L} \prod_{j \neq i}^{L} f(w_k - \lambda_j)} \prod_{j=2}^{L} \frac{\dd w_j}{2 \ii \pi} \; ,
\>
where $f'(0)$ denotes the derivative of $f(\lambda)$ with respect to $\lambda$ at the point $\lambda=0$.
The expression (\ref{ansatz1}) decomposes similarly to (\ref{red1}) allowing us to look for a term by term
identification. Thus taking into account the explicit form of the factors $m_i$ given in (\ref{mi}), we find the following relation
for the function $H$,
\<
\label{HH}
\left. H(w_1 , \dots , w_L) \right|_{w_1 = \lambda_i} &=& \frac{f'(0)}{\prod_{j=2}^{L} f(\mu_1 - \mu_j + \gamma)} \frac{f(\theta + \gamma + \mu_1 - \lambda_i)}{f(\theta + \gamma)} \prod_{j=2}^{L} f(\lambda_i - \mu_j + \gamma) \times \nonumber \\
&& \prod_{j \neq i}^{L} f(\mu_1 - \lambda_j) \prod_{j \neq i}^{L} f(\lambda_j - \lambda_i + \gamma) \prod_{j = 2}^{L} f(w_j - \lambda_i) \bar{H} (w_2 , \dots , w_L) \; . \nonumber \\
\>
The function $\bar{H}$ in (\ref{HH}) consists of $H$, up to an overall multiplicative factor independent of $\lambda_j$, $w_j$ and $\theta$, 
under the mappings $\theta \rightarrow \theta + \gamma$ and $\mu_i \rightarrow \mu_{i+1}$ \footnote{Strictly speaking, the identity (\ref{HH}) is only required to hold when integrated as
$\oint \dots \oint \left[ \qquad    \right] \dd w_2 \dots \dd w_L$.}. Furthermore, the LHS of (\ref{HH}) consists of the function
$H$ computed at the particular point $w_1 = \lambda_i$, and it would be useful to have a similar relation valid for  
general values of the variable $w_1$. In order to obtain such relation, we first notice that (\ref{HH}) needs
to be satisfied for $i \in [1,L]$ and that it is required to hold only when integrated according to (\ref{ansatz1}). 
Thus assuming that $H$ has no poles inside the integration contour, we only need to consider (\ref{HH})
under the mappings $\lambda_i \rightarrow w_1$ and $\lambda_j \rightarrow w_j$ for $j \neq i$ 
to obtain the relation
\<
\label{HHH}
H(w_1 , \dots , w_L) &=& \frac{f'(0)}{\prod_{j=2}^{L} f(\mu_1 - \mu_j + \gamma)} \frac{f(\theta + \gamma + \mu_1 - w_1)}{f(\theta + \gamma)} \prod_{j=2}^{L} f(w_1 - \mu_j + \gamma) \times \nonumber \\
&& \prod_{j \neq 1}^{L} f(\mu_1 - w_j) \prod_{j \neq 1}^{L} f(w_j - w_1 + \gamma) \prod_{j = 2}^{L} f(w_j - w_1) \bar{H} (w_2 , \dots , w_L) \; . \nonumber \\
\>
Now the formula (\ref{HHH}) can be readily iterated once we know $H(w_1)$. For that we consider the
results of the \Appref{sec:sol1} and from (\ref{T1}) we can immediately read
\[
\label{h1}
H( w_1 ) = f'(0) f(\gamma) \frac{f(\theta + \gamma - w_1 + \mu_1)}{f(\theta + \gamma)} \; .
\]
Thus the iteration of (\ref{HHH}) with (\ref{h1}) as initial condition yields the formula
\<
\label{zint}
H (w_1 , \dots , w_L) =&& [f'(0) f(\gamma)]^L  \prod_{j>i}^{L} f(w_j - w_i + \gamma) f(w_j - w_i) \prod_{j=1}^{L} \frac{f(\theta + j \gamma - w_{j} + \mu_j)}{f(\theta + j \gamma)} \times \nonumber \\
&& \prod_{j<i}^{L}  f(\mu_j - w_i) \prod_{j>i}^{L}  f(w_i - \mu_j + \gamma) \; .
\>
The expression (\ref{zint}) already takes into account the asymptotic behaviour (\ref{asy}) and 
though here we have considered a functional equation different from the one obtained
in \cite{Galleas_2011}, the expression (\ref{zint}) indeed reduces to the formula of \cite{Galleas_2011}
in the degenerated limit with the conventions properly adjusted.
Moreover, it is worth remarking that the homogeneous limit $\lambda_j \rightarrow \lambda$ and $\mu_j \rightarrow \mu$ 
can be trivially obtained from the integral formula (\ref{ansatz}, \ref{zint}).

\section{Concluding remarks}
\label{sec:conclusion}

In this work the partition function of the elliptic SOS model with domain
wall boundaries was studied through a fusion of algebraic and functional
techniques. The partition function of the model was shown to obey a functional equation
arising from commutation rules encoded in the dynamical Yang-Baxter relation (\ref{DYB})
which is valid for general values of the model parameters. The solution was then obtained as a multiple
contour integral.

The possibility of deriving functional equations for such partition functions
from the Yang-Baxter algebra and its dynamical counterpart was 
firstly demonstrated in \cite{Galleas10, Galleas_2011}. Although here
we have also employed the dynamical Yang-Baxter relation, the mechanism considered
in \Secref{sec:fun} differs from the one used in \cite{Galleas10, Galleas_2011}, and the resulting
functional equation is significantly simpler than the ones previously obtained. 
Interestingly, solving this new type of functional equation follows the same lines of \cite{Galleas_2011}
but each one of the steps required are dramatically simplified.

The elliptic SOS model considered here is also referred to as 8VSOS model
in the literature and for the special value of the anisotropy parameter $\gamma = \frac{2 \ii \pi}{3}$, 
it reduces to the so called Three-colouring model \cite{Baxter_1970}. For the case
with domain wall boundaries, the partition function of the Three-colouring model
was shown to obey a certain functional equation in \cite{Razumov_2009a, Razumov_2009b}
but a possible connection with our results has eluded us so far. 
In the work \cite{Rosengren_2011} this same partition function was studied under the light of the
symmetric polynomials theory  where a set of two-variables polynomials have been introduced. 
These polynomials were conjectured in \cite{Mangazeev_2010} to satisfy a certain partial differential
equation and recurrence relation, to which (\ref{FZ}), (\ref{ansatz}) and (\ref{zint}) might shed some light into their proofs.

\section{Acknowledgements}
\label{sec:ack}
The author thanks the Australian Research Council (ARC) and the Centre of Excellence for the 
Mathematics and Statistics of Complex Systems (\mbox{MASCOS}) for financial support.
The author also thanks the anonymous referee for several comments and suggestions which
helped to improve this manuscript.

\appendix

\section{Asymptotic behaviour}
\label{sec:asymp}

In the limit $\theta \rightarrow \infty$ the $\mathcal{R}$-matrix (\ref{rmat})
resembles the one associated with the six-vertex model, except that the Boltzmann
weights (\ref{bw}) still consist of elliptic theta-functions.
Also, from the definition (\ref{pft}) we can readily see that the whole dependence of
$Z_{\theta}$ with a particular variable $\lambda_j$ will be described by the operator
$B(\lambda_j, \theta + j\gamma)$, which is very similar to the six-vertex model
analogous in the mentioned limit. Moreover, in order to proceed with the analysis of
$Z_{\theta}$ in the full limit $(\lambda_j , \theta) \rightarrow \infty$, it will be useful
to rewrite (\ref{abcd}) as
\[
\label{abcdL}
\mathcal{T}_{a}^{(L)} (\lambda, \theta) = \left( \begin{matrix}
A_{L} (\lambda, \theta) & B_{L} (\lambda, \theta) \cr
C_{L} (\lambda, \theta) & D_{L} (\lambda, \theta) \end{matrix} \right) \; .
\]
The Eq. (\ref{abcdL}) differs from (\ref{abcd}) by the index $L$ that we have inserted
in order to emphasise we are considering the ordered product of $L$ matrices $\mathcal{R}_{a j}$ as given by (\ref{mono}).
The matrix $\mathcal{R}_{a j}(\lambda , \theta)$ in its turn consists of a $2 \times 2$
matrix in the space $\mathbb{V}_a$, i.e.
\[
\label{raj}
\mathcal{R}_{a j}(\lambda , \theta) = \left( \begin{matrix}
\alpha_j (\lambda , \theta) & \beta_j (\lambda , \theta) \cr
\gamma_j (\lambda , \theta) & \delta_j (\lambda , \theta) \end{matrix} \right) \; ,
\]
whose entries are then matrices acting non-trivially on the $j$-th space of the tensor
product $\mathbb{V}_1 \otimes \dots \otimes \mathbb{V}_L$. More precisely we have
\begin{align}
\label{alfa}
\alpha_j (\lambda , \theta) &= \left( \begin{matrix}
a_{+} (\lambda , \theta) & 0 \cr
0 & b_{+} (\lambda , \theta)  \end{matrix} \right)_j &
\beta_j (\lambda , \theta) &= \left( \begin{matrix}
0 & 0 \cr
c_{+} (\lambda , \theta) & 0  \end{matrix} \right)_j \nonumber \\
\gamma_j (\lambda , \theta) &= \left( \begin{matrix}
0 & c_{-} (\lambda , \theta) \cr
0 & 0  \end{matrix} \right)_j &
\delta_j (\lambda , \theta) &= \left( \begin{matrix}
b_{-} (\lambda , \theta) & 0 \cr
0 & a_{-} (\lambda , \theta) \end{matrix} \right)_j \; .
\end{align}
In this way the definition (\ref{mono}) can be implemented 
recursively, i.e. 
\[
\label{recur}
\mathcal{T}_{a}^{(L+1)} (\lambda, \theta) = \mathcal{T}_{a}^{(L)} (\lambda, \theta) \mathcal{R}_{a L+1} (\lambda - \mu_{L+1} , \hat{\theta}_{L+1} ) \; ,
\]
with initial conditions
\begin{align}
\label{init}
A_1 (\lambda, \theta) &= \alpha_1 (\lambda - \mu_1 , \hat{\theta}_1) & B_1 (\lambda, \theta) &= \beta_1 (\lambda - \mu_1 , \hat{\theta}_1) \nonumber \\
C_1 (\lambda, \theta) &= \gamma_1 (\lambda - \mu_1 , \hat{\theta}_1) & D_1 (\lambda, \theta) &= \delta_1 (\lambda - \mu_1 , \hat{\theta}_1) \; .
\end{align}
In particular, from (\ref{recur}) we can single out the relation
\[
\label{recurB}
B_{L+1} (\lambda , \theta) = A_L (\lambda , \theta) \beta_{L+1}(\lambda - \mu_{L+1} , \hat{\theta}_{L+1}) +  B_L (\lambda , \theta) \delta_{L+1}(\lambda - \mu_{L+1} , \hat{\theta}_{L+1}) \; ,
\]
which allows us to obtain the behaviour of $B(\lambda_j, \theta + j\gamma)$ from the analysis of
$\alpha_j$, $\beta_j$, $\gamma_j$ and $\delta_j$ in the limit $(\lambda_j , \theta) \rightarrow \infty$.
Thus taking into account (\ref{ff}), (\ref{bw}) and (\ref{alfa}), in the limit $(\lambda , \theta) \rightarrow \infty$ we find
\begin{align}
\label{alfalim}
\alpha_j &\sim \frac{1}{2} \sum_{n_j = - \infty}^{+ \infty} (-1)^{n_j - \frac{1}{2}} p_{n_j} q_{n_j}^{\frac{1}{2}} e_{n_j}^{\lambda} K_{n_j} &
\beta_j &\sim \frac{1}{2} \sum_{n_j = - \infty}^{+ \infty} (-1)^{n_j - \frac{1}{2}} p_{n_j} q_{n_j} X^{-} \nonumber \\
\gamma_j &\sim \frac{1}{2} \sum_{n_j = - \infty}^{+ \infty} (-1)^{n_j - \frac{1}{2}} p_{n_j} q_{n_j} X^{+} &
\delta_j &\sim \frac{1}{2} \sum_{n_j = - \infty}^{+ \infty} (-1)^{n_j - \frac{1}{2}} p_{n_j} q_{n_j}^{\frac{1}{2}} e_{n_j}^{\lambda} K_{n_j}^{-1} \; , \nonumber \\
\end{align}
where we have introduced the conventions $e_n = e^{-(2n + 1)}$, $p_n = p^{(n + \frac{1}{2})^2}$ and $q_n = e_n^{\gamma}$.
In their turn the operators $K_n$ and $X^{\pm}$ appearing in (\ref{alfalim}) are given by
\<
\label{gg}
K_n = \left(\begin{matrix}
q_n^{\frac{1}{2}} & 0 \\
0 & q_n^{-\frac{1}{2}} \end{matrix}
\right)
\qquad \qquad
X^{\pm} = \frac{1}{2} \left(\begin{matrix}
0 & 1 \pm 1 \\
1 \mp 1 & 0 \end{matrix}
\right) \; . 
\>

Now the relation (\ref{recurB}) can be iterated with the help of (\ref{init}) and (\ref{alfalim}). 
Thus in the limit $(\lambda_j , \theta) \rightarrow \infty$ we find the expression
\<
\label{expans}
B(\lambda_j, \theta + j\gamma) &\sim& \frac{f(\gamma)}{2^{L-1}}  \sum_{n_1 = - \infty}^{\infty} \dots \sum_{n_{L-1} = - \infty}^{\infty} (-1)^{\sum_{i=1}^{L-1} n_i - \frac{(L-1)}{2}} \prod_{i=1}^{L-1} p_{n_i} q_{n_i}^{\frac{1}{2}} e_{n_i}^{\lambda_j} \times \nonumber \\
&& \qquad \qquad \qquad \qquad \qquad \sum_{j=1}^{L} e_{n_1}^{-\mu_1} \dots e_{n_{j-1}}^{-\mu_{j-1}} P_j^{\vec{n}} e_{n_{j}}^{-\mu_{j+1}} \dots e_{n_{L-1}}^{-\mu_{L}} \; , \nonumber \\
\>
with $\vec{n} = (n_1 , \dots , n_{L-1})$ and operators $P_j^{\vec{n}}$ reading
\[
\label{pj}
P_j^{\vec{n}} = K_{n_1} \otimes \dots \otimes K_{n_{j-1}} \otimes X^{-} \otimes K_{n_{j+1}}^{-1} \otimes \dots \otimes K_{n_{L-1}}^{-1} \; .
\]
The operators $K_n$ and $X^{\pm}$ satisfy the following analogous of
the $q$-deformed $\alg{su}_2$ algebra
\<
\label{su2}
 K_n X^{\pm} K_m^{-1} &=& q_{\frac{n+m}{2}}^{\pm} X^{\pm} \nonumber \\
\left[ X^{+} , X^{-} \right] &=& \frac{K_n K_m - K_n^{-1} K_m^{-1}}{(q_{\frac{n+m}{2}} - q_{\frac{n+m}{2}}^{-1})} \; ,
\>
which allows us to demonstrate the properties
\<
\label{ppp}
P_i^{\vec{n}^{(a)}}  P_j^{\vec{n}^{(b)}} &=& q_{n_i^{(b)}} q_{n_j^{(a)}} P_j^{\vec{n}^{(b)}} P_i^{\vec{n}^{(a)}} \qquad \qquad (i < j) \nonumber \\
P_i^{\vec{n}^{(a)}} P_i^{\vec{n}^{(b)}} &=& 0 \; .
\>
As we shall see, the relations (\ref{ppp}) will be of utility for the analysis of the asymptotic behaviour of $Z_{\theta}$. 

Next, in order to analyse the behaviour of $Z_{\theta}$ in the proposed limit, we substitute the expansion
(\ref{expans}) into the definition (\ref{pft}) and use the relations (\ref{ppp}) to reorganise the result properly. By doing so we obtain
the expression
\<
\label{expans1}
&& Z_{\theta} (\lambda_1 , \dots , \lambda_L) \sim \nonumber \\
&& \frac{f(\gamma)^L}{2^{L(L-1)}} \sum_{n_1^{(1)} = - \infty}^{\infty} \dots \sum_{n_{L-1}^{(1)} = - \infty}^{\infty} \dots \sum_{n_1^{(L)} = - \infty}^{\infty} \dots \sum_{n_{L-1}^{(L)} = - \infty}^{\infty} (-1)^{\sum_{a=1}^{L} \sum_{i=1}^{L-1} n_i^{(a)} - \frac{L(L-1)}{2}} \nonumber \\
&& \qquad \qquad \qquad \prod_{a=1}^{L} \prod_{i=1}^{L-1} p_{n_i^{(a)}} q_{n_i^{(a)}}^{\frac{1}{2}} e_{n_i^{(a)}}^{\lambda_a - \mu_{i}^{(a)}} 
\; \sum_{\sigma \in \mathcal{S}_L} \prod_{(a,b) \in I_{\sigma}} (q_{n_{b-1}^{(a)}} q_{n_{a}^{(b)}})^{-1} \;  \bra{\bar{0}} \prod_{a=1}^{L} P_{a}^{\vec{n}^{(a)}} \ket{0} \; , \nonumber \\
\>
where $\mu^{(a)} = \{ \mu_i : i \neq a \}$. As usual $\mathcal{S}_L$ denotes the group of permutations of $L$
objects while $\sigma(a)$ stands for the permutation of the $a$-th object. In order to clarify the meaning of $I_{\sigma}$
let us consider the usual two row representation of $\sigma$. We draw a line starting at the object $a$ in the top row and ending in the bottom row 
at the position $\sigma(a)$ such that only two lines intersect at any one point. 
The lines are labelled by their numbers in the top row and the points of intersection are called inversion vertices.
In this way an inversion vertex can be labelled by a pair $(a,b)$ with $a<b$ such that $a$ and $b$ label the 
two intersecting lines originating the inversion vertex. Then denoting $[L] = \{1 , \dots , L \}$, we call 
$I_{\sigma} = \{ (a,b) \in [L] \times [L]: a < b \; \; \mbox{and} \; \; \sigma(a) > \sigma(b) \}$ 
the set of inversion vertices labels of a given permutation $\sigma$.

The next step to obtain an explicit expression for (\ref{expans1}) is to compute the quantity
$\bra{\bar{0}} \prod_{a=1}^{L} P_{a}^{\vec{n}^{(a)}} \ket{0}$ which can be readily performed since the operators
$P_{a}^{\vec{n}^{(a)}}$ consist of a simple tensor product (\ref{pj}). Thus considering (\ref{states1}) we obtain
\[
\bra{\bar{0}} \prod_{a=1}^{L} P_{a}^{\vec{n}^{(a)}} \ket{0} = \prod_{a=1}^{L} \prod_{i=1}^{L-1} q_{n_{i}^{(a)}}^{\frac{1}{2}} 
\] 
which can be substituted in (\ref{expans1}) yielding the formula
\<
\label{expans2}
&& Z_{\theta} (\lambda_1 , \dots , \lambda_L) \sim \nonumber \\
&& \frac{f(\gamma)^L}{2^{L(L-1)}} \sum_{n_1^{(1)} = - \infty}^{\infty} \dots \sum_{n_{L-1}^{(1)} = - \infty}^{\infty} \dots \sum_{n_1^{(L)} = - \infty}^{\infty} \dots \sum_{n_{L-1}^{(L)} = - \infty}^{\infty} (-1)^{\sum_{a=1}^{L} \sum_{i=1}^{L-1} n_i^{(a)} - \frac{L(L-1)}{2}} \nonumber \\
&& \qquad \qquad \qquad \qquad \qquad \qquad  \prod_{a=1}^{L} \prod_{i=1}^{L-1} p_{n_i^{(a)}} q_{n_i^{(a)}} e_{n_i^{(a)}}^{\lambda_a - \mu_{i}^{(a)}} 
\; \sum_{\sigma \in \mathcal{S}_L} \prod_{(a,b) \in I_{\sigma}} (q_{n_{b-1}^{(a)}} q_{n_{a}^{(b)}})^{-1} \nonumber \\
\>
in the limit $(\lambda_j , \theta) \rightarrow \infty$.

\section{Theta-function properties}
\label{sec:elliptic}

In this appendix we recall some useful properties of elliptic theta-functions that 
we have considered through this paper. We remark here that many of these properties
have also been discussed in \cite{Rosengren_2008}. The function $f$ defined in \Secref{sec:dyn}
consists basically of the Jacobi theta-function $\Theta_1$ \cite{Watson} and in this paper
we have omitted the dependence of $f$ with the elliptic nome $p$ for brevity. In what follows
we summarise some properties of elliptic theta-functions adjusted to our conventions.

\paragraph{Addition rule.} The function $f$ satisfy the addition rule
\<
\label{add}
f(\lambda_1 + \lambda_2) f(\lambda_1 - \lambda_2) f(\lambda_3 + \lambda_4) f(\lambda_3 - \lambda_4) &=& \nonumber \\
f(\lambda_1 + \lambda_4) f(\lambda_1 - \lambda_4) f(\lambda_3 + \lambda_2) f(\lambda_3 - \lambda_2) &+& f(\lambda_1 + \lambda_3) f(\lambda_1 - \lambda_3) f(\lambda_2 + \lambda_4) f(\lambda_2 - \lambda_4) \; . \nonumber \\
\> 
 
\paragraph{Analyticity and periodicity.} The function $f$ is an entire function, that is to say all
of its singularities are removable, and it has only simple zeroes. It is also an odd function and quasi doubly-periodic, i.e.
\[
\label{dp}
f(\lambda - \ii \pi) = - f(\lambda) \qquad \qquad \qquad  f(\lambda - \ii \pi \tau) = - e^{2 \lambda - \ii \pi \tau} f(\lambda) \; .
\]

\paragraph{Trigonometric limit.} In the limit $p \rightarrow 0$ the theta-function $f(\lambda)$
degenerate into a trigonometric function. More precisely we have $\lim_{p \to 0} - \ii p^{-\frac{1}{4}} f(\lambda) = \sinh(\lambda)$,
which allows for an easy comparison with previous results in the literature.

\paragraph{Higher order theta-functions.} For a fixed value of the elliptic nome $\tau$, we call $\mathcal{F}$ 
a theta-function of order $L$ and norm $t$ if 
\[
\label{fff}
\mathcal{F}(\lambda) = C \prod_{j=1}^{L} f(\lambda - \chi_j)  
\]
for constants $C$ and $\chi_j$ such that $\sum_{j=1}^{L} \chi_j = t$. Moreover, due to (\ref{dp}) one can
readily show the quasi-periodicity
\<
\label{DP}
\mathcal{F}(\lambda - \ii \pi) &=& (-1)^L  \mathcal{F} (\lambda) \nonumber \\   
\mathcal{F}(\lambda - \ii \pi \tau) &=& (-1)^L e^{2 (L \lambda - t) - \ii \pi \tau L} \mathcal{F}(\lambda) \; .
\>
In fact, the factorised form (\ref{fff}) and the quasi-periodicity (\ref{DP}) for entire functions can be shown to 
be equivalent properties \cite{Weber}. This feature allows us to state a more general result.   
Let $\bar{\mathcal{F}}$ be defined as
\[
\label{ffff}
\bar{\mathcal{F}}(\lambda) = \sum_i \bar{C}_i \prod_{j=1}^{L} f(\lambda - \chi_j^{(i)})  \; ,
\]
with $\bar{C}_i$ being constants and $\sum_{j=1}^{L} \chi_j^{(i)} = t$ for any $i$. The function $\bar{\mathcal{F}}$
is entire and obeys the quasi-periodicity (\ref{ffff}), thus it can be factored similarly
to (\ref{fff}).

\paragraph{$Z_{\theta}$ as a higher order theta-function.} The partition function $Z_{\theta}$
defined in (\ref{pft}) is written as a product of operators $B(\lambda, \theta)$. As a matter of fact, the
whole dependence of $Z_{\theta}$ with a particular variable $\lambda_j$ is contained in a single operator 
$B(\lambda_j, \theta + j\gamma)$ since the vectors $\ket{0}$ and $\ket{\bar{0}}$ are constants. 
Now taking into account (\ref{bw}), (\ref{alfa}) and (\ref{init}), the recurrence relation (\ref{recurB})
tells us that the entries of $B(\lambda_j, \theta + j\gamma)$ are of the form (\ref{fff}). This is because 
the factors $A_L (\lambda , \theta) \beta_{L+1}(\lambda - \mu_{L+1} , \hat{\theta}_{L+1})$
and $B_L (\lambda , \theta) \delta_{L+1}(\lambda - \mu_{L+1} , \hat{\theta}_{L+1})$ in (\ref{recurB})
do not contribute simultaneously to the same entry due to the structure of (\ref{alfa}). Thus, due to the definition (\ref{pft}), we have that
the partition function $Z_{\theta}$ will be of the form (\ref{ffff}) with respect to a given variable $\lambda_j$.
The latter characterises $Z_{\theta}$ as a higher order theta-function of order $L$ in each one of the variables $\lambda_j$.
Although its explicit value shall not be required through this work, we shall use $t_j$ to denote the norm of $Z_{\theta}$ when
factored with respect to the variable $\lambda_j$ as given by (\ref{fff}).

\section{$Z_{\theta}$ as a symmetric function}
\label{sec:symmetricfun}

The commutativity of operators $B(\lambda, \theta)$ as described by (\ref{alg}),
together with the definition (\ref{pft}), implies that $Z_{\theta}$ is a symmetric function.
This commutativity has been extensively employed in the derivation of (\ref{FZ})
and here we intend to show that this symmetry becomes an inherent property of the solutions of
(\ref{FZ}). 

Through the inspection of the coefficients (\ref{coeff}), we notice that $N_i \leftrightarrow N_j$
under the mapping $\lambda_i \leftrightarrow \lambda_j$ while $M_0 \rightarrow M_0$ and $N_k \rightarrow N_k$ for $k \neq i,j$.
Thus performing this mapping on (\ref{FZ}) and subtracting it from the original equation we obtain the following relation,
\<
\label{prv}
&& M_0 [ Z_{\theta - \gamma} (\lambda_1 ,\dots, \lambda_i , \dots , \lambda_j , \dots , \lambda_L) - Z_{\theta - \gamma} (\lambda_1 ,\dots, \lambda_j , \dots , \lambda_i , \dots , \lambda_L)] \nonumber \\
&+& N_0 [ Z_{\theta} (\lambda_1 ,\dots, \lambda_i , \dots , \lambda_j , \dots , \lambda_L) - Z_{\theta} (\lambda_1 ,\dots, \lambda_j , \dots , \lambda_i , \dots , \lambda_L)] \nonumber \\  
&+& \sum_{k=1}^{L} N_k  Z_{\theta} (\lambda_0 ,\dots, \lambda_i , \dots , \lambda_{k-1}, \lambda_{k+1} , \dots , \lambda_j , \dots)  \nonumber \\
&-& \sum_{k=1}^{L} N_k  Z_{\theta} (\lambda_0 ,\dots, \lambda_j , \dots , \lambda_{k-1}, \lambda_{k+1} , \dots , \lambda_i , \dots)  = 0 \; .
\>
Next we solve (\ref{prv}) for the $l$-th term of the summation over the index $k$ which yields the expression
\<
\label{prf}
&& \frac{M_0}{N_l} [ Z_{\theta - \gamma} (\lambda_1 ,\dots, \lambda_i , \dots , \lambda_j , \dots , \lambda_L) - Z_{\theta - \gamma} (\lambda_1 ,\dots, \lambda_j , \dots , \lambda_i , \dots , \lambda_L)] \nonumber \\
&+& \frac{N_0}{N_l} [ Z_{\theta} (\lambda_1 ,\dots, \lambda_i , \dots , \lambda_j , \dots , \lambda_L) - Z_{\theta} (\lambda_1 ,\dots, \lambda_j , \dots , \lambda_i , \dots , \lambda_L)] \nonumber \\  
&+& \sum_{\stackrel{k=1}{k \neq l}}^{L} \frac{N_k}{N_l}  Z_{\theta} (\lambda_0 ,\dots, \lambda_i , \dots , \lambda_{k-1}, \lambda_{k+1} , \dots , \lambda_j , \dots)  \nonumber \\
&-& \sum_{\stackrel{k=1}{k \neq l}}^{L} \frac{N_k}{N_l}  Z_{\theta} (\lambda_0 ,\dots, \lambda_j , \dots , \lambda_{k-1}, \lambda_{k+1} , \dots , \lambda_i , \dots)  =  \nonumber \\
&&Z_{\theta} (\lambda_0 ,\dots, \lambda_j , \dots , \lambda_{l-1}, \lambda_{l+1} , \dots , \lambda_i , \dots) - Z_{\theta} (\lambda_0 ,\dots, \lambda_i , \dots , \lambda_{l-1}, \lambda_{l+1} , \dots , \lambda_j , \dots) \; . \nonumber \\
\>
The RHS of (\ref{prf}) does not depend on $\lambda_l$ so this variable can be chosen such that the LHS of (\ref{prf}) vanishes. 
Thus we can conclude that
\[
Z_{\theta} (\lambda_0 ,\dots, \lambda_j , \dots , \lambda_{l-1}, \lambda_{l+1} , \dots , \lambda_i , \dots) = Z_{\theta} (\lambda_0 ,\dots, \lambda_i , \dots , \lambda_{l-1}, \lambda_{l+1} , \dots , \lambda_j , \dots) \; ,
\]
and since this is valid for any $i$, $j$ and $l$, the symmetry property
\[
\label{symmetry}
Z_{\theta} (\dots , \lambda_i , \dots , \lambda_j , \dots) = Z_{\theta} (\dots , \lambda_j , \dots , \lambda_i , \dots)
\]
immediately follows.

\section{Solution for $L=1$}
\label{sec:sol1}

The functional equation (\ref{FZ}) for $L=1$ explicitly reads
\<
\label{FZ1}
M_0 Z_{\theta - \gamma} (\lambda_1) + N_0 Z_{\theta} (\lambda_1) + N_1 Z_{\theta} (\lambda_0) = 0
\>
with coefficients
\<
\label{C1}
M_0 &=& \frac{f(\theta)}{f(\theta + \gamma)} f(\lambda_0 - \mu_1) \nonumber \\
N_0 &=& - \frac{f(\theta + \gamma)}{f(\theta + 2\gamma)} f(\lambda_0 - \mu_1 + \gamma) \frac{f(\lambda_1 - \lambda_0 + \gamma)}{f(\lambda_1 - \lambda_0)} \nonumber \\
N_1 &=& \frac{f(\theta + \gamma + \lambda_0 - \lambda_1)}{f(\theta + 2\gamma)} \frac{f(\gamma)}{f(\lambda_1 - \lambda_0)} f(\lambda_1 - \mu_1 + \gamma) \; .
\>
By setting $\lambda_0 = \lambda_1 - \theta - \gamma$, the coefficient $N_1$ vanishes and (\ref{FZ1}) simplifies to
\[
\label{fzt}
Z_{\theta} (\lambda_1) \frac{f(\theta + \gamma)}{f(\theta + \gamma + \mu_1 - \lambda_1)} = Z_{\theta - \gamma} (\lambda_1) \frac{f(\theta)}{f(\theta + \mu_1 - \lambda_1)} \; .
\]
The relation (\ref{fzt}) is an equation only over the variable $\theta$ which is readily solved by
\[
\label{tt}
Z_{\theta} (\lambda_1) = \frac{f( \theta + \gamma - \lambda_1 + \mu_1)}{f(\theta + \gamma)} F(\lambda_1) 
\]
where $F$ is $\theta$ independent. After eliminating the dependence with $\theta$, we can substitute (\ref{tt})
back into (\ref{FZ1}). The resulting equation can then be simplified and we obtain the relation
\[
\label{ttt}
\frac{f(\gamma) f(\theta + \gamma + \mu_1 - \lambda_0) f(\theta + \gamma + \lambda_0 - \lambda_1) f(\lambda_1 - \mu_1 + \gamma)}{f(\theta + \gamma) f(\theta + 2\gamma) f(\lambda_1 - \lambda_0)}
(F(\lambda_1) - F(\lambda_0)) = 0 \; .
\]
From (\ref{ttt}) we can conclude that $F$ is a constant and it can be fixed by
the asymptotic behaviour (\ref{expans2}). Thus we find for $L=1$,
\[
\label{T1}
Z_{\theta} (\lambda) = f(\gamma) \frac{f(\theta + \gamma - \lambda + \mu_1)}{f(\theta + \gamma)}  \; .
\]

\bibliographystyle{hunsrt}
\bibliography{references}

\end{document}